\documentclass[journal]{IEEEtran}

\usepackage[utf8]{inputenc} 
\usepackage{amsmath, amssymb, amsfonts} 
\usepackage{graphicx} 
\usepackage{tikz} 
\usepackage{cite} 
\usepackage{hyperref} 
\usepackage{xcolor} 
\usepackage{algorithm}
\usepackage{algorithmic}
\usepackage{microtype}
\usepackage{pgfplots}
\usepackage{physics}
\usepackage{subcaption}
\usepackage{booktabs}
\usepackage{listings}
\usepackage{xcolor}
\usepackage{tcolorbox}
\definecolor{darkgray}{rgb}{0.15,0.15,0.15}  
\usepackage{csquotes}
\usepackage[normalem]{ulem}

\definecolor{codegray}{rgb}{0.5,0.5,0.5}
\lstdefinestyle{mystyle}{
    language=Python,
    basicstyle=\ttfamily\footnotesize,
    keywordstyle=\color{blue}\bfseries,
    commentstyle=\color{gray},
    stringstyle=\color{red},
    numbers=left,
    numbersep=5pt,
    numberstyle=\tiny\color{codegray},
    frame=lines,
    breaklines=true
}

\pgfplotsset{compat=1.18}

\usetikzlibrary{arrows, shapes, arrows.meta, automata, positioning, shapes.geometric, graphs, backgrounds, external,  fit, calc, shadows, patterns, patterns.meta}
\begin{document}

\title{Maestro: Intelligent Execution for Quantum Circuit Simulation}
\author{
    Oriol Bertomeu\textsuperscript{1}, Hamzah Ghayas\textsuperscript{1}, Adrian Roman\textsuperscript{1},  and $^*$Stephen DiAdamo\textsuperscript{1} \\[1ex]
    \textsuperscript{1}Qoro Quantum, Munich, Germany \\     
    $^*$stephen@qoroquantum.de    
}

\maketitle

\maketitle

\begin{abstract}

Quantum circuit simulation remains essential for developing and validating quantum algorithms, especially as current quantum hardware is limited in scale and quality. However, the growing diversity of simulation methods and software tools creates a high barrier to selecting the most suitable backend for a given circuit. We introduce Maestro, a unified interface for quantum circuit simulation that integrates multiple simulation paradigms—state vector, MPS, tensor network, stabilizer, GPU-accelerated, and p-block methods—under a single API. Maestro includes a predictive runtime model that automatically selects the optimal simulator based on circuit structure and available hardware, and applies backend-specific optimizations such as multiprocessing, GPU execution, and improved sampling. Benchmarks across heterogeneous workloads demonstrate that Maestro outperforms individual simulators in both single-circuit and large batched settings, particularly in high-performance computing environments. Maestro provides a scalable, extensible platform for quantum algorithm research, hybrid quantum-classical workflows, and emerging distributed quantum computing architectures.

\end{abstract}

\begin{IEEEkeywords}
    Quantum computing, high performance computing, quantum circuit simulation, quantum emulation, GPU simulation, quantum software.
\end{IEEEkeywords}

\section{Introduction}
Quantum circuit simulation plays a central role in the development and validation of quantum algorithms, particularly given the limited scale and noise levels of current hardware. A wide range of simulators exist today—general-purpose libraries such as Qiskit Aer\cite{qiskit}, Cirq\cite{cirq}, and ProjectQ\cite{projectq}; hardware-optimized tools like NVIDIA cuQuantum\cite{cuquantum}; distributed-computing-focused simulators such as QuEST\cite{quest}; and commercial platforms including AWS Braket and Azure Quantum. These systems employ different strategies, including state vector, stabilizer, matrix product state (MPS), and tensor-network methods, often with optional GPU acceleration.

While this ecosystem provides powerful capabilities, it also creates a substantial barrier for practitioners. Each simulator exposes its own API, configuration options, and performance trade-offs, requiring users to manually tune circuits, adapt them to specific backends, and reason about the interaction between circuit structure and available hardware. This fragmentation complicates hybrid quantum-classical workflows and limits scalability, particularly in high-performance or batch-computing environments where circuit heterogeneity makes fixed pipelines inefficient. HPC centers and cloud providers face similar challenges: maintaining diverse simulators with different dependencies and hardware assumptions introduces operational overhead that scales poorly as new tools emerge.

To address these challenges, we present Maestro, a unified interface for quantum circuit simulation. Maestro compiles circuits into a common intermediate representation, translates them into the native formats of multiple backend simulators, and automatically applies backend-specific optimizations such as multithreading, multiprocessing, and efficient sampling. A central feature is Maestro’s dynamic backend-selection engine, which analyzes circuit features—such as gate density, entanglement structure, and measurement layout—to predict the fastest simulator for each circuit. This eliminates much of the manual tuning typically required and ensures more consistent throughput across heterogeneous workloads.

Maestro further introduces performance optimizations around existing simulators, including GPU-accelerated state vector and MPS backends, improving execution for large or highly structured circuits. By abstracting away simulator-specific complexity while exploiting backend-specific strengths, Maestro simplifies simulation for researchers and developers and provides a scalable deployment layer for HPC centers and cloud providers. In anticipation of emerging distributed quantum computing models, Maestro also supports p-block simulation, enabling early benchmarking and evaluation of multi-node quantum architectures.

Beyond its technical contributions, Maestro is designed with practical deployment considerations in mind. As quantum computing becomes increasingly integrated into scientific workflows, HPC centers and service providers must support multiple simulators with heterogeneous requirements. Maestro mitigates this burden by offering a unified orchestration layer through which diverse simulation backends can be deployed, managed, and exposed to users.

The core features of Maestro are open-sourced under the GNU General Public License version 3\cite{maestro_repo}, with additional optional capabilities—such as the automatic backend-selection engine and GPU-based execution—provided through secondary dynamically loaded libraries under a commercial license.

In this paper, we describe the design and implementation of Maestro (Section~\ref{sec:system}) and present experimental benchmarks across a diverse set of circuits and backends (Section~\ref{sec:bench}). Our results demonstrate that Maestro simplifies simulator usage and deployment while improving performance across heterogeneous, high-performance workloads. We further validate Maestro in an HPC settings through experiments and integration with three HPC environments, illustrating its applicability for large-scale, batch-oriented simulation environments. We conclude with broader applications and future directions in Section~\ref{sec:conclusion}.

\section{Simulation Methods and Frameworks}\label{sec:simulation}

Classical simulation remains essential for developing and validating quantum algorithms, especially given the limits of current hardware. Yet simulation costs grow rapidly with qubit count and entanglement, motivating a range of techniques that balance accuracy, memory usage, and runtime. Below, we summarize the main approaches relevant to Maestro and highlight why automated backend selection is necessary.

State vector simulation represents the full quantum state as a vector of size $2^n$, permitting exact results and full-state inspection. Its drawback is exponential memory growth, which makes simulations above roughly 30 qubits impractical without specialized hardware. Despite this, state vector methods remain ideal for small- to medium-scale circuits requiring precision.

Partitioned-block (p-block) simulation alleviates single-node memory limits by allocating one simulator per qubit and merging them only when interactions occur~\cite{viamontes2009quantum}. This dynamic growth enables circuits with localized entanglement to run where a monolithic state vector would exceed memory. When entanglement is global, p-block degenerates to standard state vector simulation, but it offers significant advantages for distributed or measurement-heavy circuits.

MPS simulators exploit low entanglement by encoding states as tensor chains with bounded bond dimension~\cite{vidal2003efficient, schollwock2011density, orus2014practical}. This drastically reduces memory needs for shallow or structured circuits. However, entanglement growth increases the bond dimension exponentially, making MPS unsuitable for highly entangled workloads.

Tensor network simulation generalizes MPS to arbitrary network geometries~\cite{orus2014practical, schollwock2011density}. These methods scale well for circuits with limited entanglement or regular connectivity, such as error-correcting codes or many-body systems, but become expensive when entanglement spreads widely.

GPU-accelerated simulators~\cite{qiskit2024, bayraktar2023cuquantum} parallelize matrix operations across thousands of cores and can provide large speedups for wide circuits. Their performance, however, depends heavily on memory bandwidth and transfer overheads, and GPU simulation may be slower than optimized CPU methods at smaller scales or for batched workloads~\cite{faj2023quantum}.

Each method involves trade-offs in accuracy, scalability, and resource efficiency. state vector approaches excel at exactness; MPS and tensor networks handle low-entanglement circuits; GPU backends accelerate large tensor contractions; and p-block methods support distributed or communication-heavy circuits. Hardware characteristics (CPU cores, memory, GPU availability) further complicate performance prediction, making optimal backend selection difficult for users—particularly in heterogeneous batch workloads.

Shot execution and expectation-value computations adds additional complexity. Simulation cost depends on both single-run performance and the efficiency with which a simulator handles repeated executions. Inefficient shot handling can dominate runtime, especially for applications requiring thousands of samples.

Simulator ecosystems vary widely in interfaces and features~\cite{jamadagni2024benchmarking}. For example, Qiskit Aer supports multiple simulation modes and performs some automatic selection~\cite{qiskit2024}, but does not choose among MPS or other advanced methods, leaving users to make expert decisions.

Given this diversity, a unified and automated simulation framework is needed to leverage the strengths of each method without requiring users to understand low-level trade-offs. Maestro provides such an interface by integrating multiple simulation paradigms, automating backend selection, and optimizing execution across CPU, GPU, and distributed settings. The design is detailed in the next section.

\section{Maestro: Software Design and Key Features}\label{sec:system}
Maestro is designed as a modular and extensible platform that unifies diverse quantum circuit simulators under a single umbrella interface. Its architecture is organized around five key components: (1) a backend abstraction layer to map commonly used circuit representations (e.g., QASM) to native circuit data structures of simulators (2) integrations with a broad set of simulator backends, including those that are able to run on GPUs (3) parallelization strategies leveraging multiprocessing and multithreading, (4) optimizations for efficient sampling of measurement outcomes, and (5) a prediction engine that selects the most appropriate simulator backend for a given circuit. Together, these components provide a consistent interface for end-users, enable simple addition of new backends, and ensure that both algorithm developers and HPC providers can efficiently execute large and heterogeneous simulation workloads.

In this section, we detail the traits of Maestro that make it easier to use, richer in features, and often more efficient than other commonly used simulators. First, we explore how one can access many different simulation methods through a single interface. Then, we detail how Maestro makes efficient use of the computing resources available with multiprocessing and multithreading. Next, we explain how our backend prediction method works to estimate the fastest performing backend. Finally we explain our GPU-based implementation.

\subsection{Backend Abstraction}

A central design objective of Maestro is to provide a unified abstraction layer that decouples users from simulator-specific interfaces while preserving native execution performance. To this end, Maestro defines an intermediate circuit representation that enables seamless compilation to a wide range of simulation backends. This design allows users to execute the same quantum program across different simulators without manual translation or modification of their circuits.

At present, Maestro integrates simulators based on state vector, stabilizer, MPS, tensor network, and p-block methods. Each backend is encapsulated through a standardized adapter that maps Maestro’s intermediate circuit representation onto the simulator’s native API. This modular architecture facilitates the incorporation of new backends with minimal engineering effort, while isolating simulator-specific dependencies from the core system.

\begin{figure*}[t]
    \centering
    \includegraphics[]{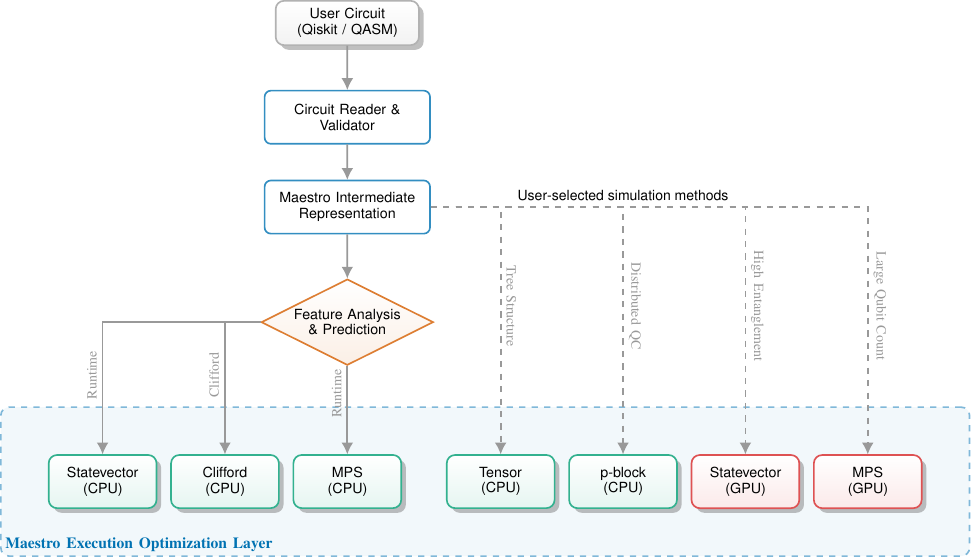}
    \caption{Architectural overview of the Maestro interface. User circuits are ingested into a common Intermediate Representation (IR). The Prediction Engine analyzes circuit features (e.g., entanglement entropy, gate density) to dynamically route execution to the optimal simulation backend. }
    \label{fig:maestro_architecture}
\end{figure*}

Currently, Maestro accepts Qiskit circuits and OpenQASM2 as input. Leveraging Qiskit’s extensive interoperability, many external circuit formats can be converted into Qiskit prior to execution. Once a circuit is imported, Maestro transforms it into its internal representation and subsequently reconstructs it in the native syntax of the selected simulator. This process preserves backend-specific optimizations and ensures that each simulator operates at its full native performance.

Beyond serving as a unifying abstraction, Maestro introduces cross-simulator performance enhancements. These include optimized sampling routines, advanced threading configurations, and execution scheduling strategies that are applied independently of the backend. As demonstrated in the benchmarking section, these optimizations frequently result in superior performance compared to running the simulators directly, highlighting the benefits of Maestro’s layered architecture.

\subsection{Multiprocessing and Multithreading}

To efficiently exploit modern computing architectures, Maestro is designed to automatically use the full computational capacity of the host node. The system dynamically adapts to the available hardware by distributing computation across accessible CPU cores and threads.

The execution strategy is determined by the circuit's structure and the chosen backend. Initially, the circuit is executed on a single simulator instance up to the first reset or measurement; this instance may utilize internal multithreading if supported (e.g., state vector simulations exceeding a defined size threshold). For the sampling phase, the parallelization logic diverges based on the measurement topology:

\begin{itemize}
    \item If \textbf{mid-circuit measurements or resets} are present, the simulator state is cloned, and execution proceeds via multiple single-threaded instances running concurrently across available threads.
    \item If \textbf{measurements occur only at the end}, state vector simulators utilize optimized sampling without thread-level cloning. However, other backends (such as MPS) retain the multithreaded cloning approach to efficiently manage state saving and restoring.
\end{itemize}

Maestro can also compute expectation values with comparable cost to a single-shot evaluation, leveraging cached circuit states.

Notably, GPU-backed simulators are exempt from this CPU-level threading logic, strictly maintaining a single execution thread to leverage the GPU's intrinsic parallel architecture.

This multiprocessing paradigm is naturally compatible with HPC environments, where compute nodes can be allocated dynamically. Maestro's parallel execution model allows simulations to scale seamlessly across HPC clusters, efficiently exploiting multi-core and multi-node architectures for large-scale or distributed quantum workloads.

\subsection{Backend Prediction}

Maestro incorporates a predictive model for runtime estimation and backend selection to automatically determine the most efficient simulator for a given quantum circuit. This prediction is based on a multi-regression model trained on empirical performance data collected from a large set of benchmark circuits. The model employs fine-grained timing analysis, in which numerous circuit instances are executed while recording detailed gate-level statistics, for each simulator type. For MPS simulations, we also vary the bond dimension. These measurements are combined with known algorithmic complexity estimates of each simulation method to predict the expected computational cost for new circuits.

Maestro's prediction engine avoids ``black box'' learning in favor of a component-wise algorithmic summation model. We define the total execution time $T_{total}$ as the sum of discrete operation costs, where each operation is modeled as the product of a theoretical complexity term ($O_{op}$) and a hardware-calibrated coefficient ($C_{op}$):
\begin{equation}
    T_{total} \approx \sum_{op \in \mathcal{C}}  C_{op}(\mathbf{x}) \cdot O_{op}(\mathbf{x}) 
\end{equation}
where $\mathbf{x}$ represents the circuit and simulation features. The complexity classes $O_{op}$ differ fundamentally by backend:

\begin{enumerate}
    \item \textit{State-Vector Estimation:} For state-vector simulation, complexity is dominated by the Hilbert space size $O(2^n)$. However, memory bandwidth and threading efficiency introduce non-linearities. Maestro handles this by benchmarking operations across varying qubit counts ($n$) to generate a univariate coefficient curve $C_{op}(n)$. At runtime, we estimate specific gate costs using piecewise cubic Hermite interpolation. For multi-shot execution, we apply a linear regression model $T_{shots}(S) \approx c_1 + c_2 \cdot S$ to account for sampling overhead. We use this sampling approach because we employ alias sampling, which has a $O(1)$ sampling overhead~\cite{walker1974new}.

    \item \textit{Matrix Product State Estimation:} The MPS estimator uses a more sophisticated bivariate model dependent on both qubit count ($n$) and bond dimension ($\chi$). Analyzing the tensor contraction costs yields two distinct complexity classes:
    \begin{itemize}
        \item \textbf{Local Updates:} Single-qubit gates involve localized tensor contractions scaling as $O(\chi^2)$.
        \item \textbf{Entanglement \& Measurement:} Two-qubit gates and measurements are dominated by Singular Value Decomposition (SVD) and the overhead of maintaining canonical forms across the chain. We model these operations with the scaling term $O(n \cdot \chi^3)$.
    \end{itemize}
    To capture hardware performance for these operations, Maestro maintains a calibration surface $C_{op}(n, \chi)$ constructed from benchmarks on a grid of $(n, \chi)$ pairs. During prediction, we employ bivariate Hermite interpolation on this surface to determine the exact pre-factor for the theoretical $O(n \cdot \chi^3)$ complexity.
\end{enumerate}

To ensure hardware independence, the coefficient surfaces ($C_{op}$) are not static. A one-time benchmark routine can run during installation, executing synthetic circuits to capture the host machine's specific FLOPs, memory bandwidth, and thread scaling factors $\alpha(t)$. This ensures that a prediction made on a laptop versus an HPC node reflects the local hardware capabilities, while the underlying algorithmic complexity terms ($O_{op}$) remain constant.

During execution, Maestro applies this model in real time to predict the runtime across all available simulators and automatically select the backend expected to deliver the fastest execution. This process also includes structural circuit analysis; for example, when circuits composed entirely of Clifford gates are being processed, stabilizer simulators are taken into consideration. In this case, Maestro will make a runtime estimate against the selected simulators and stabilizer simulators, because other simulators may still run faster.

\begin{figure}[ht]
    \centering    
    \includegraphics[]{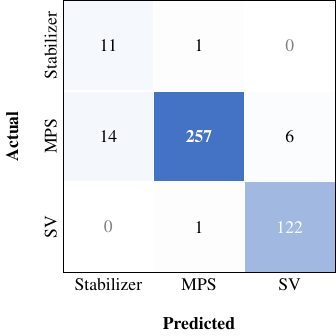}
    \caption{Predicted vs actual optimal simulators confusion matrix. To account for possible variance in execution time due to the run time environment in cases where multiple simulators are of the same quality, we choose to validate simulations where Maestro was off by less than $0.01$ s, with that error being no more than $10\%$ of the optimal runtime. }
    \label{fig:confusion-matrix}
\end{figure}

A confusion matrix can be seen in Fig.~\ref{fig:confusion-matrix}, where we classified 412 circuits against the true runtime. The classification model demonstrates strong overall performance with a global accuracy of 95\%, excelling particularly in distinguishing MPS and SV instances, both of which achieve F1-scores above 0.96. However, the stabilizer class reveals a distinct performance asymmetry. While the model is highly sensitive to true stabilizers (Recall: 0.92), it suffers from low precision (0.44) due to a specific confusion pattern: 14 MPS states were incorrectly classified as stabilizer. This suggests that while the model rarely misses a true stabilizer, it is prone to false positives, often mistaking certain MPS for stabilizers. These false positives occur primarily in low-entanglement circuits where the runtime delta between MPS and Stabilizer is negligible, resulting in minimal impact on total batch duration. Nevertheless, as we expand on in Sec.~\ref{fig:batch-circuits}, the simulation instances where this confusion pattern appears, namely Clifford circuits with low entanglement, are also characterized by short run times on the order of the milliseconds, which means the overall performance of the prediction layer does not see significant degradation.

\begin{table}[h]
\centering
\begin{tabular}{lcccc}
        \toprule
        \textbf{Class} & \textbf{Precision} & \textbf{Recall} & \textbf{F1-Score} & \textbf{Support} \\ 
        \midrule
        Stabilizer     & 0.44               & 0.92            & 0.59              & 12 \\
        MPS            & 0.99               & 0.93            & 0.96              & 277 \\
        State vector    & 0.95               & 0.99            & 0.97              & 123 \\ 
        \midrule
        \textbf{Accuracy} & & & \textbf{0.95} & 412 \\
        \textbf{Macro Avg} & 0.79 & 0.95 & 0.84 & 412 \\
        \bottomrule
    \end{tabular}
    \caption{Accuracy scores for Maestro's prediction capability.}
\end{table}

Currently, Maestro predicts the best CPU-based simulations from its supported state vector, Matrix Product State (MPS), and stabilizer simulators (specifically for Clifford circuits). However, accurate prediction for GPU-based simulation remains a distinct challenge. The primary difficulty is that the performance ratio between CPU and GPU execution varies significantly depending on the specific circuit structure, rendering generalized benchmarks unreliable. Consequently, while general trends exist, we have not yet incorporated GPU performance prediction into our model. This is an immediate next step for future work.

In addition, we have explored alternative machine learning models, including decision tree and neural network-based predictors. Preliminary results indicate that while these models can improve prediction accuracy, they are generally highly hardware-dependent and less portable. Ongoing work focuses on combining these approaches into a hybrid model that balances predictive precision with generalizability across hardware platforms.

\subsection{GPU Simulation}

To enable GPU-accelerated quantum circuit simulation while preserving the unified interface of Maestro, the framework repurposes its backend abstraction layer for both CPU and GPU execution paths. For example, state-vector simulations are mapped onto the cuStateVec library of the NVIDIA cuQuantum SDK~\cite{bayraktar2023cuquantum}.

For the MPS method, however, we found that the existing MPS support in cuQuantum (via cuTensorNet) did not offer all required features—most notably, access to separate singular values within the factorised representation. Therefore, we developed a custom MPS backend over cuQuantum to complement Maestro’s abstraction layer. This allows users to continue using Maestro’s circuit input and switching logic, while transparently activating GPU-accelerated execution paths.

Consequently, Maestro users benefit from accessible GPU-simulator execution without changing their circuit description: the same Qiskit-derived, or QASM2, circuit can be compiled and dispatched either to CPU or GPU backends under Maestro’s control. This design ensures that performance gains from GPU hardware are achieved with minimal friction for the user.

\subsection{Distributed Quantum Computing}

Distributed quantum computing (DQC) involves partitioning quantum circuits to match a specific network topology and injecting communication instructions to produce a logically equivalent distributed circuit. These communication steps rely on entanglement distribution and classical communication, where the latter consists of information bits derived from qubit measurements.

To simulate a networked cluster of quantum computers, one can either house the entire simulation in a single monolithic memory block or utilize p-block simulation to manage memory dynamically. The p-block approach offers a decisive advantage here. For example, simulating a network of five quantum computers, each with 20 qubits, would require a state vector capable of handling 100 qubits if pre-allocated monolithically—an impossibility for any existing supercomputer. Conversely, p-block simulation initializes with low memory requirements and dynamically resizes the simulator instances only when entanglement demands it.

In DQC architectures, interconnectivity is managed via specific communication qubits. Under p-block simulation, these qubits can be isolated in their own memory partitions. When a link is established, the communication qubits interact, temporarily merging into the active simulation state. Crucially, once the communication protocol (such as teleportation) concludes—typically via measurement or reset—these communication qubits are effectively removed from the active simulation context.

While the \textit{computational} qubits of two interacting hosts may remain fully entangled (requiring a combined simulation state), this approach avoids the overhead of maintaining persistent state for communication qubits. Unlike a monolithic simulator, which must statically allocate memory for the communication qubits alongside the hosts, the p-block strategy treats them as transient resources. By discarding these two qubits from the entangled state immediately after use (rendering them independent), the required state vector size remains at least $4\times$ smaller (a reduction of the Hilbert space by a factor of $2^2$) compared to a monolithic equivalent that permanently includes the communication infrastructure. The trade-off is that DQC circuits rely heavily on mid-circuit measurements, which introduce synchronization overhead that can impact overall execution time.

To produce distributed quantum circuits, we have developed an automated toolchain allowing users to submit standard monolithic circuits (e.g., via Qiskit or OpenQASM) which are then automatically optimized for distributed execution. The system uses graph partitioning algorithms to remap qubits and minimize the ``cut size''—the number of non-local gates requiring expensive network communication. By reordering qubits to maximize locality within each memory block, the system reduces latency overhead while maintaining exact simulation precision.

This architecture enables Maestro to support advanced simulation requirements that are often incompatible with standard fragmentation techniques. Specifically, the DQC mode supports:
\begin{itemize}
    \item \textbf{Automated Circuit Remapping:} Dynamic translation of monolithic circuits into distributed, topology-aware sub-circuits.
    \item \textbf{Mid-Circuit Measurements:} Full support for dynamic circuits where measurement results influence control flow, synchronized across distributed partitions.
    \item \textbf{State vector Optimization:} The p-block architecture is implemented  for CPU-based state vector simulators. 
\end{itemize}

By breaking the hard memory ceiling of single-node architectures, Maestro enables the verification of large-scale distributed algorithms—potentially scaling to 100+ qubits in low-entanglement scenarios—providing a vital verification tool for future modular quantum architectures.

\section{Performance Benchmarking}\label{sec:bench}
To evaluate Maestro’s performance and validate that its abstraction layer preserves simulator quality, we adopt a structured benchmarking methodology. Our benchmarking is organized into four dimensions: (i) general benchmarking for correctness and baseline runtime, (ii) GPU acceleration, (iii) batched circuit execution, and (iv) distributed quantum computing using p-block simulators. Across all experiments, we focus on two key metrics: fidelity and runtime. Fidelity is assessed first to ensure the correctness of simulation outputs; only simulators achieving a minimum threshold are considered for runtime comparisons. The circuits we use for these benchmarks come from the dataset found in \cite{leonteva2025comparative}.

\subsection{General Benchmarking}

A suite of circuits was constructed to capture representative algorithmic patterns, including GHZ states (entanglement generation), quantum Fourier transform (interference), random Clifford+T circuits (universal gate sets), and QAOA layers of increasing depth (variational workloads). All circuits terminate in final measurements to remove any simulator-dependent variability in mid-circuit measurement handling.  

Runtime benchmarks are then conducted on circuits of varying width and depth under consistent conditions, such as fixed shot counts and hardware allocations. This allows us to characterize not only absolute performance but also scaling behavior with problem size. 

To assess the benefits of GPU acceleration, we compare CPU and GPU backends across both state vector and matrix product state (MPS) simulations. We report times-to-solution for both CPU and GPU simulations, and identify break-even points where GPU acceleration begins to outperform CPU execution. 

\begin{figure*}[h!]
    \centering

    \begin{subfigure}[b]{0.3\textwidth}
        \centering
        \resizebox{\linewidth}{!}{\includegraphics[]{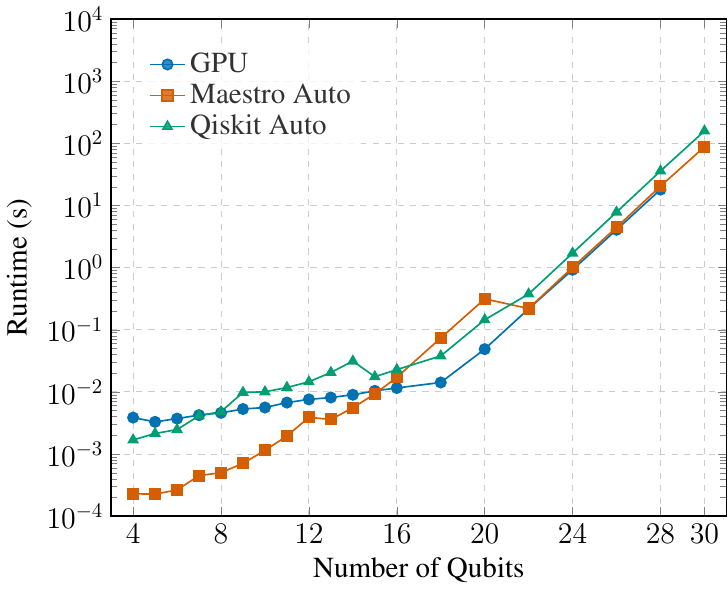}}        
        \caption{Amplitude Estimation (AE)}
        \label{fig:statevector-gpu-ae}
    \end{subfigure}
    \hfill
    \begin{subfigure}[b]{0.3\textwidth}
        \centering
        \resizebox{\linewidth}{!}{\includegraphics[]{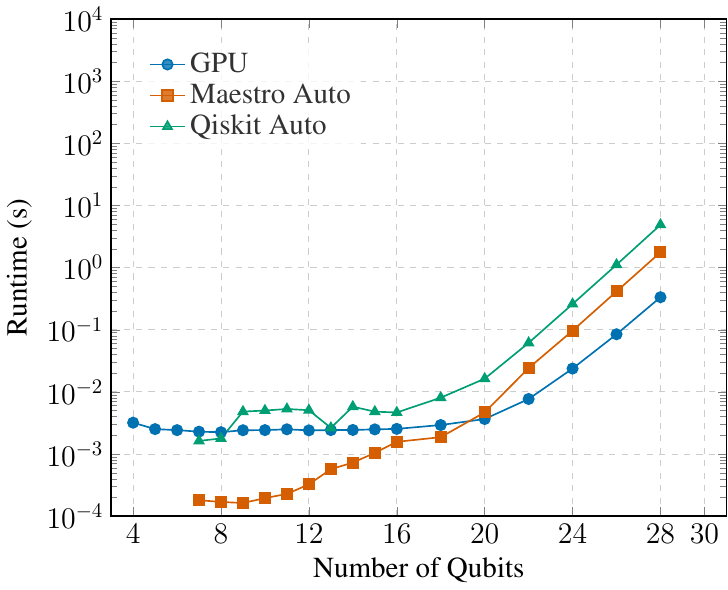}}
        \caption{GHZ state generation}
        \label{fig:statevector-gpu-ghz}
    \end{subfigure}
    \hfill
    \begin{subfigure}[b]{0.3\textwidth}
        \centering
        \resizebox{\linewidth}{!}{\includegraphics[]{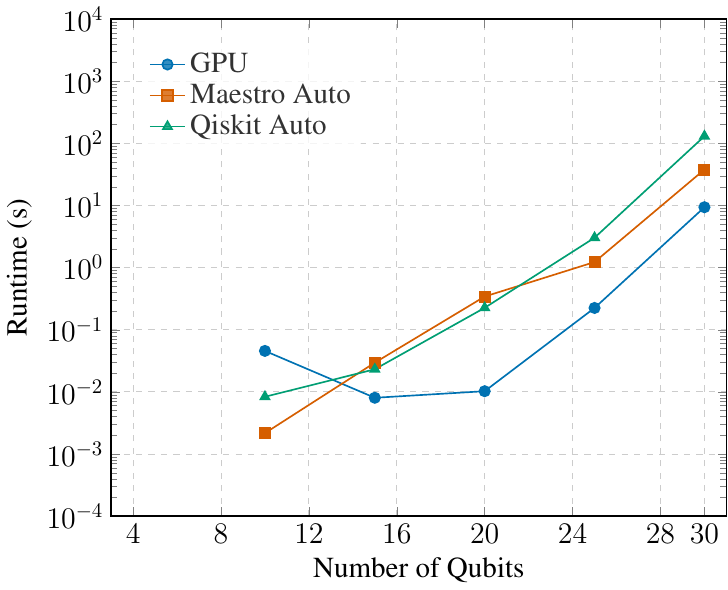}}
        \caption{QAOA}
        \label{fig:statevector-gpu-qaoa}
    \end{subfigure}

    \vspace{0.5cm}

    \begin{subfigure}[b]{0.3\textwidth}
        \centering
        \resizebox{\linewidth}{!}{\includegraphics[]{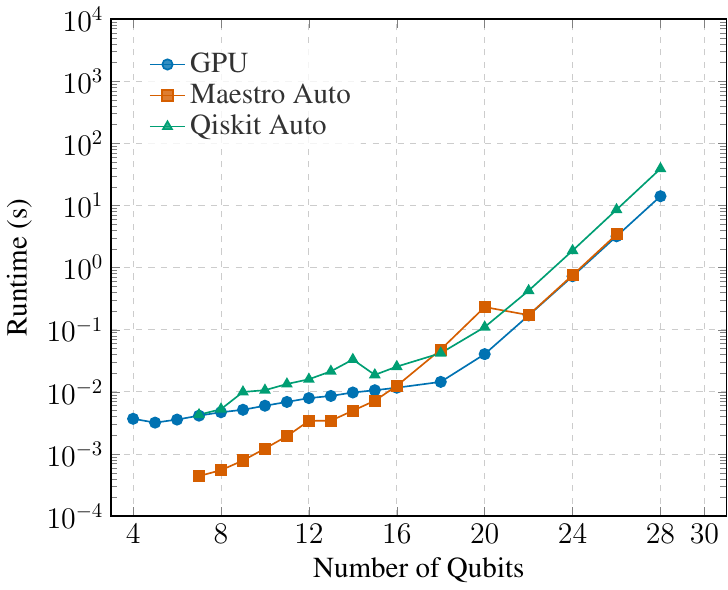}}
        \caption{Entangled quantum Fourier transform}
        \label{fig:statevector-gpu-qpe}
    \end{subfigure}
    \hfill
    \begin{subfigure}[b]{0.3\textwidth}
        \centering
        \resizebox{\linewidth}{!}{\includegraphics[]{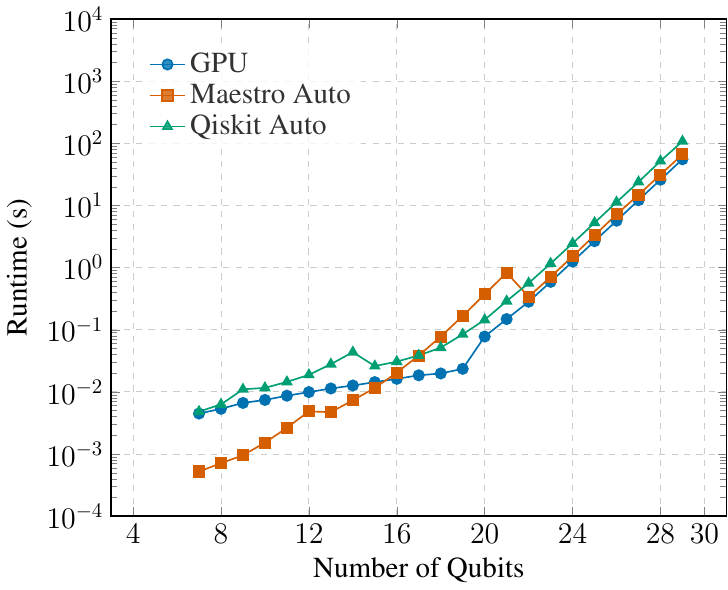}}
        \caption{Quantum neural network (QNN)}
        \label{fig:statevector-gpu-qnn}
    \end{subfigure}
    \hfill
    \begin{subfigure}[b]{0.3\textwidth}
        \centering
        \resizebox{\linewidth}{!}{\includegraphics[]{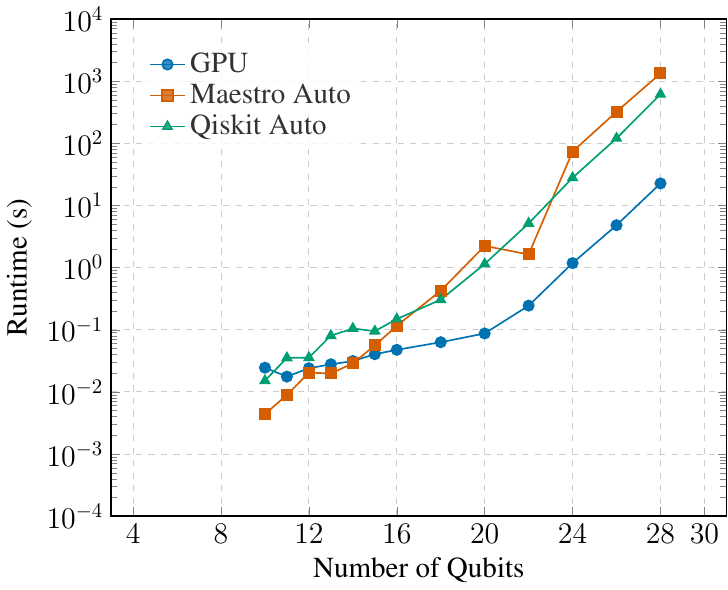}}
        \caption{Random circuit}
        \label{fig:statevector-gpu-random}
    \end{subfigure}

    \caption{Statevector-simulable task runtimes by backend, for a selection of circuit types. The GPU simulation used a NVIDIA T4 GPU and 4 vCPUs with 15 GB of memory, while the rest were run exclusively on 16 vCPUs with 124 GB of memory. 1,000 shots were captured.}
    \label{fig:statevector-gpu}
\end{figure*}

Figure \ref{fig:statevector-gpu} confirms the presence of windows of GPU advantage for state vector simulations. However, its magnitude is highly dependent on the characteristics of the circuit, and only becomes apparent when circuits are wide enough. Currently, GPU offloading is a user-configurable toggle. Future work will incorporate GPU transfer overhead into the prediction model to automate this decision.

Regarding MPS simulations, we begin by evaluating runtimes and simulator fidelity using \emph{mirror circuit fidelity}~\cite{proctor2022establishing, proctor2022measuring} as the primary measure. We have verified this approach by comparing the fidelity values it yields with fidelity values obtained through state vector amplitudes, which can be recovered for sufficiently narrow circuits, as well as by using Hellinger fidelity to compare count distributions for wider ones. We qualitatively observed that mirror fidelity tracks fidelity similarly obtained through the other estimators, and does indeed act as a good quality metric.

MPS simulations are characterized by the presence of the bond dimension ($\chi$) threshold in the tensors as a hyperparameter. While simulating with smaller tensors will tend to reduce time-to-solution, the solution will also become less accurate, due to information being lost in tensor truncation. To guarantee quality standards, we set a fidelity threshold for acceptance equal to $0.95$ and, starting from a low bond dimension threshold of four, we iteratively simulate the circuits, doubling the bond dimension parameter until the fidelity threshold is met. 


\begin{figure*}[h!] 
    \centering 

    \vspace{2mm}
    \begin{subfigure}[b]{0.3\textwidth}
        \centering
        \resizebox{\linewidth}{!}{\includegraphics[]{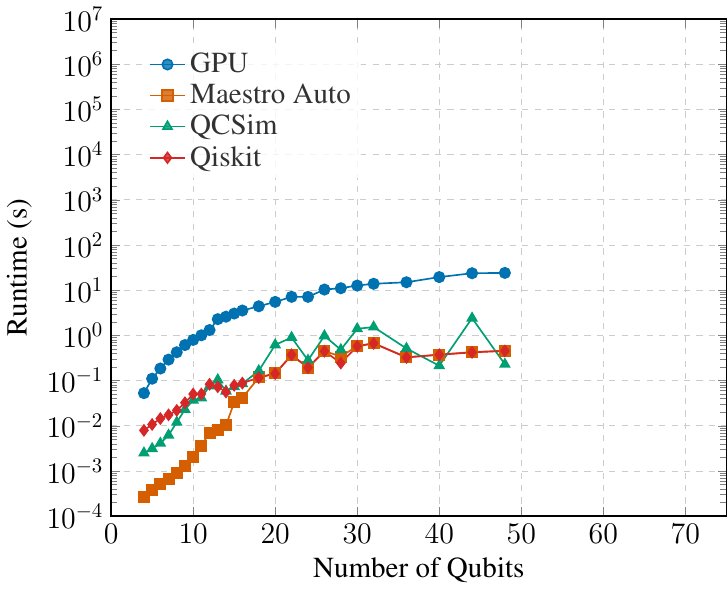}}
        \caption{Amplitude Estimation (AE)}
        \label{fig:mps-ae}
    \end{subfigure}
    \hfill 
    \begin{subfigure}[b]{0.3\textwidth}
        \centering
        \resizebox{\linewidth}{!}{\includegraphics[]{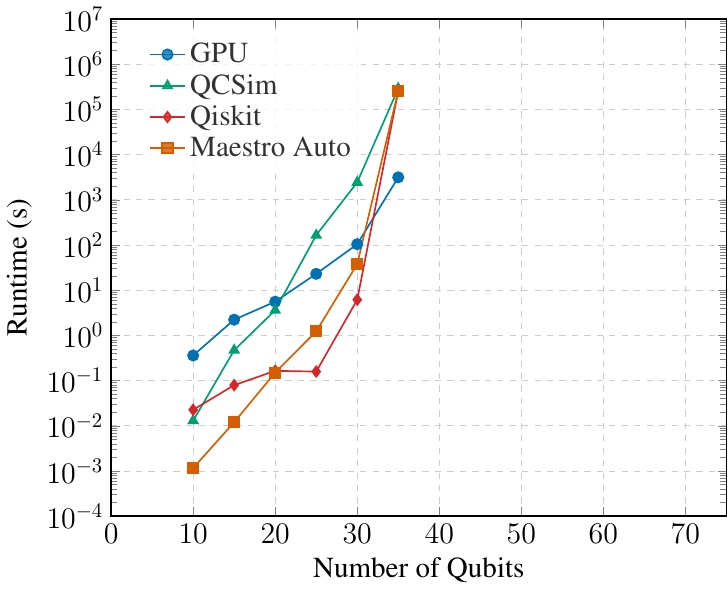}}
        \caption{QAOA ($p_{\text{link}} = 0.05$)}
        \label{fig:mps-ghz}
    \end{subfigure}
    \hfill 
    \begin{subfigure}[b]{0.3\textwidth}
        \centering
        \resizebox{\linewidth}{!}{\includegraphics[]{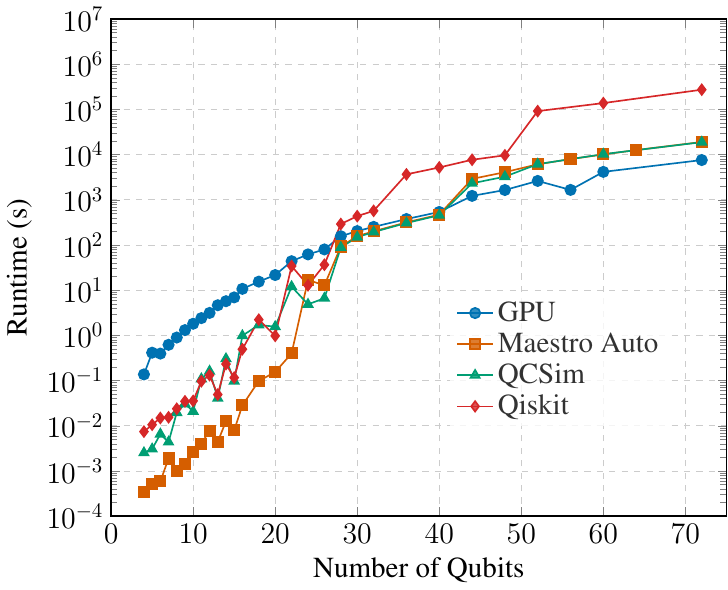}}
        \caption{Quantum Neural Network (QNN)}
        \label{fig:mps-qnn}
    \end{subfigure}
    
    \caption{Runtimes of MPS simulations for different simulators. The fidelity threshold for acceptance is set at $0.95$. For each circuit, the bond dimension is initialized at four. It is then iteratively doubled in subsequent runs until the acceptance threshold is met. Only the runtime of the final, successful iteration is recorded. 1,000 shots are taken for all simulations. }
    \label{fig:mps}
\end{figure*}

We record time-to-solution for the fastest simulation passing the fidelity threshold, and present the results in Fig.~\ref{fig:mps}. We compare multiple CPU-based simulators, including QCSim and Qiskit, as well as the automatic modes of both Maestro and Qiskit, with the GPU-based simulator. 

\subsection{Batching Circuits} \label{sec:batching-circuits}

Many practical environments, particularly HPC and cloud systems, must process large numbers of heterogeneous circuits submitted by different users. In such a multi-tenant environment, a ``one-size-fits-all" simulation strategy is inefficient: a state vector simulator wastes memory on Clifford circuits, while a tensor network simulator may stall on highly entangled instances. 

To evaluate Maestro in this setting, we constructed a ``Torture Test" batch consisting of 90 circuits designed to span the full spectrum of simulation complexity. The batch composition includes:
\begin{itemize}
    \item \textbf{Clifford circuits:} Miscellaneous circuits composed entirely of Clifford gates.
    \item \textbf{Low-entanglement circuits:} GHZ states, shallow QAOA instances ($p=1$), and hardware-efficient Ans\"{a}tze, favorable for MPS simulation.
    \item \textbf{High-entanglement circuits:} Densely entangled circuits, which require state vector simulation or high-bond-dimension MPS.
\end{itemize}

We define the \emph{Automatic Mode} as Maestro's feature-based prediction engine, which extracts circuit features (e.g., number of qubits, non-Clifford gate count, estimated entanglement entropy) prior to execution. Based on these features, the engine predicts the optimal backend (Clifford, MPS, or state vector) to minimize time-to-solution.

We compare the total batch execution time under these policies:
\begin{enumerate}
    \item \textbf{QCSim:} The QCSim simulator is used, with state vector simulations when the qubit count is less or equal to 30 qubits and MPS simulations in other cases. In the latter case, MPS bond dimension is set high enough that mirror fidelity is equal or higher than 0.95.
    \item \textbf{Qiskit:} Same as the above, but using a Qiskit-based simulator instead.
    \item \textbf{Maestro Auto:} The simulator backend is selected dynamically for each circuit based on pre-execution feature analysis.
    \item \textbf{Qiskit Auto:} The simulator backend is selected dynamically for each circuit based on Qiskit Aer's pre-execution feature analysis, falling back to Qiskit Aer's MPS simulator for circuits of more than 30 qubits, to be processed by the automatic mode.
\end{enumerate}

\begin{figure}
    \centering
    \resizebox{\linewidth}{!}{\includegraphics[]{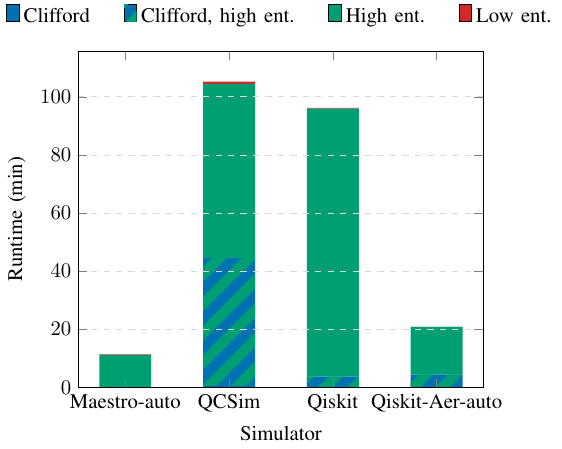}}
    \caption{Time required to execute the whole circuit batch for each of the policies. A fidelity threshold of $0.95$ was required for MPS simulations, and run times include sampling $1000$ shots per circuit.} 
    \label{fig:batch-circuits}
\end{figure}

Fig.~\ref{fig:batch-circuits} illustrates the throughput gains achieved by Maestro's automatic mode. While standard QCSim and Qiskit policies handle distinct edge cases (like very narrow or heavily entangled circuits) well, they lack the nuance to switch strategies for intermediate cases. For example, they fail to exploit MPS for suitable state-vector-sized circuits or stabilizer methods for Clifford-only circuits. Maestro mitigates this by filtering the batch: it routes computationally cheap circuits to lightweight backends and reserves heavy resources for complex tasks.

This targeted routing results in speedups of 9.2× and 8.4× over standard QCSim and Qiskit policies, respectively. Even compared to Qiskit Aer's native automatic mode, Maestro achieves a 1.8× improvement, driven by deeper optimization and broader circuit detection capabilities.

In diverse batches, highly entangled circuits typically dictate the total runtime. Maestro addresses these bottlenecks not just through efficient MPS simulation, but by matching every individual circuit to its optimal solver. This approach yields orders-of-magnitude improvements in specific cases, such as using stabilizer simulation for highly entangled Clifford circuits. Finally, the overhead required to predict the best strategy is negligible compared to the execution time, validating the efficiency of model-based batch processing.

\subsection{Distributed Quantum Computing Benchmarks}

To evaluate Maestro’s capabilities for distributed quantum computing (DQC), we perform a series of benchmarks using its \emph{p-block} simulation mode. This mode enables the execution of large-scale quantum circuits across multiple simulated QPUs connected through virtual entanglement links, thereby emulating the behavior of a distributed quantum network.
Qubits are allocated across QPUs in such a way that the number of entangling gates across QPUs is minimized \cite{AndresMartinez_Heunen2019}, \cite{kaur2025optimizedquantumcircuitpartitioning}, as these represent the bottleneck in implementing distributed quantum computation.

Maestro can be configured as a (complete) network of simulators creating a collection of logical nodes, each representing a quantum device within a fully connected topology. When executing a circuit exceeding a node's local capacity, Maestro automatically partitions the circuit and remaps qubits to other nodes, introducing the necessary inter-node gates to perform distributed quantum computing. 

The purpose of these experiments is twofold. First, they provide a systematic way to study distributed quantum computing algorithms under controlled network configurations. By simulating varying node counts and interconnect topologies, we can characterize how circuit decomposition and communication overhead influence both runtime and result fidelity. Second, these tests validate the correctness of DQC execution: by comparing distributed runs against equivalent monolithic simulations, we can quantify how closely distributed execution approximates a single coherent computation. Metrics such as the Hellinger fidelity between measurement outcomes serve as a quantitative measure of this equivalence.

In this distributed setting, the advantage of the p-block simulation lies primarily in its ability to utilize simulators of the smallest possible size. Whereas a monolithic simulation must allocate the full quantum state within a single contiguous memory space, an increasingly restrictive requirement as circuit width grows, p-block simulation partitions the system dynamically, distributing qubits across nodes only when needed. This reduces peak memory consumption and allows the simulation to scale gracefully with circuit size, since each local simulator handles only a subset of the global state.

The computational cost of managing inter-block entanglement introduces an overhead, but this can be offset by the efficiency gained from parallel execution and reduced memory contention, particularly at larger qubit counts. As a result, p-block simulation behaves similarly to real distributed quantum computing systems, where limited qubit counts per node and finite communication bandwidth determine overall performance. With this, we can scale DQC simulations up significantly, allowing a deeper study on the topic of efficient remapping methods for DQC.

We validate the results of the distributed approach to quantum computation by comparing the sampling results to those obtained through a monolithic state vector simulation. We focus on analyzing circuits that output peak states in order to minimize sampling requirements, as the number of samples required to accurately compare count distributions scales exponentially with the number of qubits in general.

\begin{figure}
    \centering
    \resizebox{\linewidth}{!}{\includegraphics[]{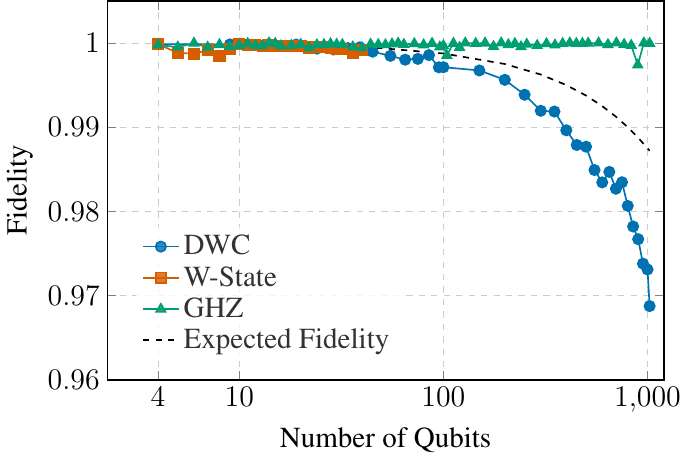}}
    \caption{DQC fidelity for circuits with known distributions. Fidelity scaling obtained for DQC simulation with a p-block state vector simulation for selected circuits. Every 20 qubits, another vQPU is added. The dashed line shows the expected fidelity result for the W-state and domain wall circuits, accounting for sampling error. }
    \label{fig:distributed-peak}
\end{figure}

Figure \ref{fig:distributed-peak} showcases the results of our implementation of simulated distributed quantum computing. In distributed quantum computing, sampling becomes more challenging and resource-consuming compared to in monolithic systems, implying a compromise between accurate fidelity reporting for verification and compute limitations. More precisely, for $S$ shots and $N$ qubits, with $S \gg N$, and a quantum state with support $O(N)$, it can be shown that the value of the Hellinger fidelity between the ideal and the sampled distribution will decrease linearly as $H \sim 1 - N/8S$. We compare our results to this bound, observing that, even for very deep (more than $1000$ qubits) circuits, about half of the fidelity defect is explained by this phenomenon.

\begin{figure*}[h!] 
    \centering 
    
    \begin{subfigure}[b]{0.3\textwidth}
        \centering
        \resizebox{\linewidth}{!}{\includegraphics[]{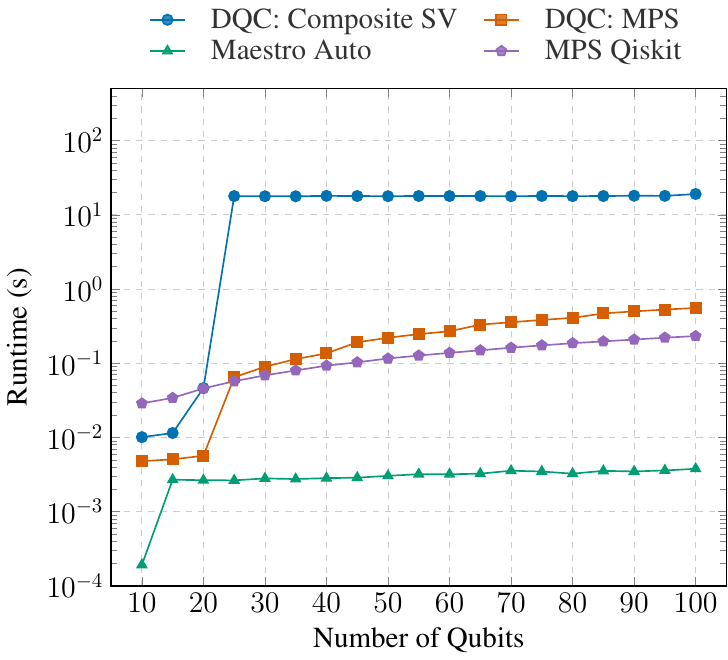}}
        \caption{GHZ state generation}
        \label{fig:mps-ae}
    \end{subfigure}
    \hfill 
    \begin{subfigure}[b]{0.3\textwidth}
        \centering
        \resizebox{\linewidth}{!}{\includegraphics[]{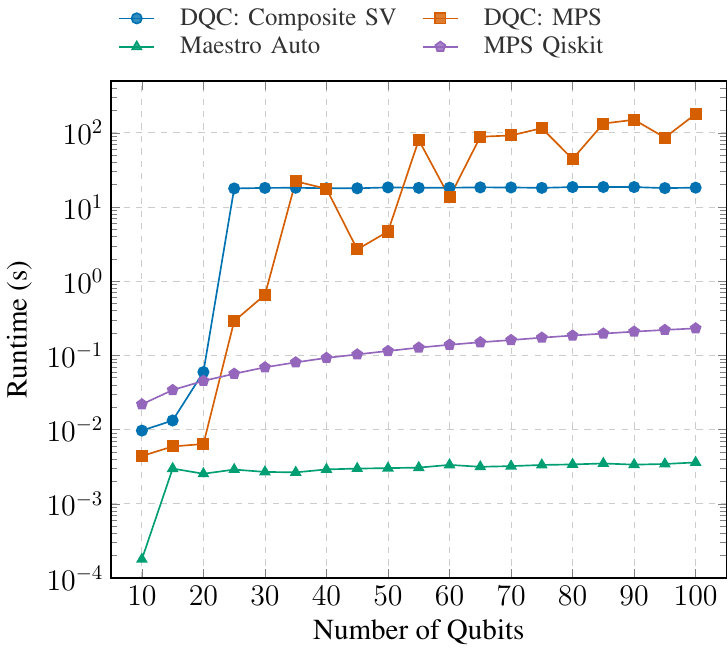}}
        \caption{Hardware-efficient ansatz (HEA)}
        \label{fig:mps-ghz}
    \end{subfigure}
    \hfill 
    \begin{subfigure}[b]{0.3\textwidth}
        \centering
        \resizebox{\linewidth}{!}{\includegraphics[]{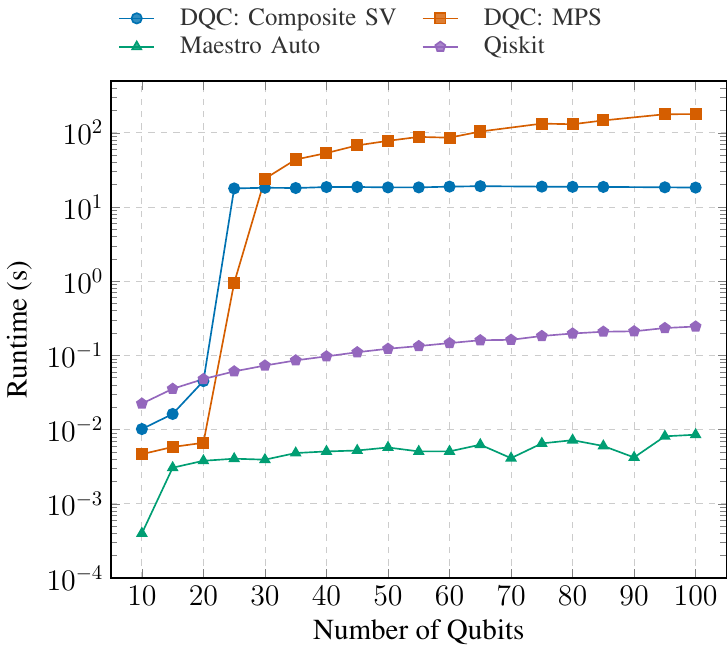}}
        \caption{Sparse QAOA}
        \label{fig:mps-qnn}
    \end{subfigure}

    \caption{Benchmarks on simulated distributed quantum computing platforms compared to results on monolithic devices. A fidelity acceptance threshold of $0.95$ was employed for all simulations. For the DQC simulations, 5 virtual QPUs were initialized with 20 qubits each.}
    
    \label{fig:distributed}
\end{figure*}

In Fig.~\ref{fig:distributed}, we present the run time performance of some distributed simulations compared to other monolithic simulations. We observe that, as expected, DQC performs like a monolithic simulation when only one vQPU is in play. However, as soon as the process becomes distributed across multiple vQPUs, we notice an immediate degradation in execution time, as the simulation becomes dominated by the cost of communication between vQPUs. Remarkably, if the appropriate simulator is used, constant run times independent of qubit depth can be observed. This allows for simulation of very deep circuits in the same time as circuits just a few tens of qubits deep. 

Our results demonstrate the possibility of using distributed quantum computing for overcoming qubit count restrictions in current devices, but they also lay bare its limitations. Namely, the fact that sharing entanglement between devices becomes a bottleneck makes simulating circuits with high entanglement challenging. We observe this when simulating W state generation, where poorly-scaling entanglement sharing needs mean simulation with more than a handful of vQPUs is not possible. To overcome this and achieve general-purpose distributed quantum computing, efficient entanglement sharing protocols have to be developed.

\section{Integration of Maestro in HPC}
A core objective of Maestro is to operate seamlessly within production-grade HPC environments. To validate its portability, stability, and ability to interface with heterogeneous compute stacks, Maestro was deployed and tested across three independent HPC environments and independently benchmarked in a scientific evaluation setting. These activities demonstrate that Maestro’s architecture can be adopted without requiring changes to existing scheduler configurations or user workflows. By leveraging each center’s established quantum software frameworks, we ensured that Maestro could be accessed through familiar HPC interfaces while still enabling advanced capabilities such as multi-backend simulation and GPU-accelerated simulators. The following subsections summarize the deployments with CESGA’s CUNQA platform, LRZ’s QDMI environment, and the independent benchmarking conducted at NPL.

\subsection{Integration with CUNQA at CESGA}

Maestro was integrated into the CUNQA platform at CESGA~\cite{vazquez2025cunqa} as an additional simulator family within CUNQA’s virtual-QPU (vQPU) architecture. CUNQA emulates distributed quantum computing (DQC) models inside an HPC environment, and its vQPUs rely on a pluggable simulator component to execute quantum tasks under the three supported communication schemes: no-communication, classical-communication, and quantum-communication. As part of our integration, we extended CUNQA’s simulator layer so that Maestro could be deployed as a vQPU backend, enabling CUNQA users to execute circuits using Maestro’s multi-backend simulation capabilities and GPU-accelerated simulators. 

Beyond adding Maestro as a simulator option, we contributed the necessary low-level mechanisms to support Maestro in CUNQA's classical-communication and quantum-communication modes. This included developing the communication-channel handling between vQPUs, allowing Maestro to participate in distributed executions that involve classical message passing or quantum-communication primitives such as teledata and telegate. Through this integration, CUNQA users can treat Maestro as a fully compatible vQPU simulator capable of executing distributed workloads across all three DQC models within the CESGA HPC environment.

\begin{figure}[h!] 
    \centering 
    
    \begin{subfigure}[b]{0.45\textwidth}
        \centering
        \resizebox{\linewidth}{!}{\includegraphics[]{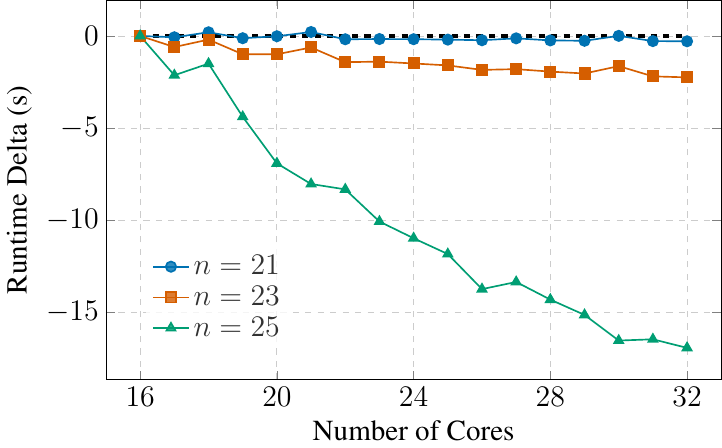}}
        \caption{Pricing calls circuits}
        \label{fig:cores_pricing}
    \end{subfigure}
        
    \begin{subfigure}[b]{0.45\textwidth}
        \centering
        \resizebox{\linewidth}{!}{\includegraphics[]{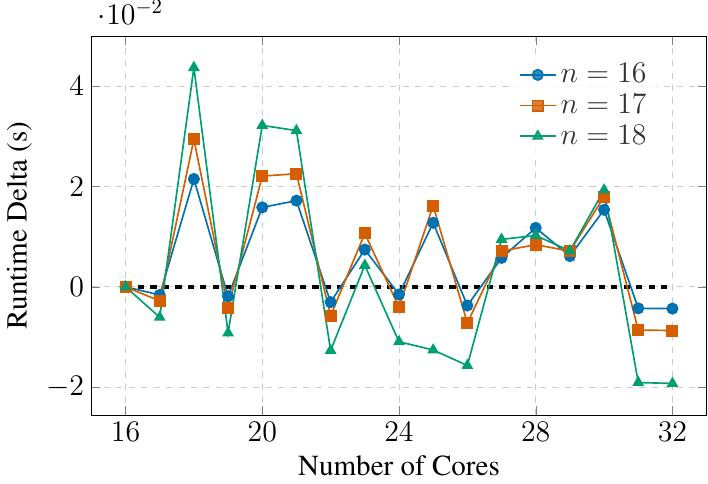}}
        \caption{Portfolio VQE circuits}
        \label{fig:cores_vqe}
    \end{subfigure}
    
    \caption{Benchmarks with varying cores compared to running with a 16-core baseline. The runtime delta is the difference in runtime compared to running with 16 cores. $n$ represents the number of qubits.}    
    \label{fig:cunqa}
\end{figure}

In Fig.~\ref{fig:cunqa}, we include a simulation benchmark that varies the number of cores used to execute various sized simulations. We can see that there are cases where performance can be vastly improved using more cores for larger sized simulations, and in some cases the performance can be more unpredictable. For example, in Fig.~\ref{fig:cores_pricing}, for pricing call benchmark circuits, as the number of cores grows, the difference in runtime clearly falls. On the other hand, in Fig.~\ref{fig:cores_vqe}, there is no clear trend. In some cases, the same circuit can take longer with more cores. This can come down to how threading is managed on the computing node, and is something we will explore more deeply in follow up benchmarks. We hypothesize that it can be a thread contention issue, and will investigate in follow-up benchmarking work.

\subsection{Integration as a QDMI at LRZ}

At the Leibniz Supercomputing Centre (LRZ), Maestro was integrated via the Quantum Device Management Interface (QDMI), the hardware–software interface of the Munich Quantum Software Stack (MQSS)~\cite{wille2024qdmi}. QDMI provides a standardized, device-agnostic interface for session management, job submission, and querying device properties, and is designed to support both physical quantum devices and virtual backends such as simulators. In our deployment, Maestro is exposed to MQSS as a QDMI-compatible backend: quantum kernels compiled by the MQSS middle-end are submitted through QDMI to a dedicated Maestro plugin, which forwards them to Maestro’s internal simulation interface. From the perspective of MQSS and LRZ users, Maestro thus appears as an additional quantum backend within the QDMI ecosystem, while internally it can exploit its multi-backend simulation capabilities, GPU-enabled simulators, and automatic backend-selection mode to execute the requested circuits efficiently.

\subsection{Benchmarks from the National Physical Laboratory}

In addition to HPC integration using platform interfaces, Maestro was independently benchmarked by the UK National Physical Laboratory (NPL) as part of the M4Q program. This external assessment provided a neutral, scientific validation of Maestro’s performance characteristics, numerical stability, and applicability to various quantum workloads. NPL evaluated Maestro across a set of representative circuits and simulation methods, comparing its runtime scaling, memory efficiency, and accuracy against established simulation tools. These benchmarks demonstrated that Maestro can deliver competitive performance while maintaining the flexibility required for heterogeneous quantum-classical workflows.

In the full report, they quote \enquote{[The] Maestro framework [is] well-suited for HPC environments due to [its] ability to exploit parallelism through multithreading and multiprocessing. Features such as Maestro Auto for batched execution and distributed simulation strategies enable efficient scaling across clusters and reduce overhead compared to single-threaded runs. Initial observations show good scalability with increased thread counts, and batch mode further improves throughput.}

Taken together, these results highlight that Maestro’s design principles translate effectively into practice, providing a scalable and robust simulation layer that supports the broader objectives of distributed quantum–classical computing.

\section{Conclusion}\label{sec:conclusion}
In this work, we presented Maestro, a unified and extensible interface for quantum circuit simulation. By abstracting over heterogeneous backends, integrating multiple simulation strategies, and introducing optimizations for parallel execution and sampling, Maestro simplifies simulator usage while consistently improving performance across diverse workloads. Our benchmarks demonstrate both high simulation fidelity and runtime efficiency, with particular advantages in GPU-accelerated, batched, and distributed scenarios. 

A key innovation of Maestro is its simulation prediction engine, which estimates expected runtimes and automatically selects the most suitable backend for a given circuit. Beyond improving user productivity and throughput in HPC environments, this capability also lays the foundation for higher-level orchestration. In particular, Maestro’s runtime estimations are leveraged within Qoro’s network layer to inform resource allocation decisions, ensuring that hybrid quantum-classical workloads can be scheduled efficiently across distributed compute resources. This positions Maestro as a practical enabler for HPC+QC integration and quantum-centric supercomputing, where quantum and classical resources must be jointly orchestrated at scale.

Looking ahead, we plan to extend Maestro in two main directions. First, by incorporating additional simulators and distributed computing execution strategies via, for example, MPI, we aim to further broaden the scope of workloads that can be executed seamlessly. Second, by deepening the integration between Maestro’s prediction engine and orchestration frameworks, we envision Maestro not only as a simulation interface but also as an enabling technology for resource-aware distributed quantum computing. In the context of HPC+QC and emerging quantum-centric supercomputers, these extensions will allow Maestro to evolve into a key software layer for managing heterogeneous, large-scale quantum-classical infrastructures.

As quantum computing continues to advance, the gap between algorithm development and practical deployment remains a critical challenge. Maestro helps bridge this gap by lowering the barrier to entry for researchers, enabling HPC centers and cloud providers to manage diverse simulators through a single interface, and preparing the ground for distributed quantum computing. In particular, Maestro’s predictive backend selection and batch-oriented execution model align with the emerging paradigm of HPC+QC integration and quantum-centric supercomputing, where quantum and classical resources must be orchestrated side by side. By combining fidelity guarantees, runtime optimizations, and predictive backend selection, Maestro supports both immediate scientific exploration and the longer-term development of scalable quantum-classical infrastructures. In doing so, it contributes not only a technical tool but also a pathway toward more accessible, efficient, and interoperable quantum computing ecosystems.

\section*{Acknowledgments}

The authors thank the National Physical Laboratory, Deep Lall and Ivan Rungger, for their support via the M4Q initiative in benchmarking and reporting on Maestro's performance which led us to make various performance improvements. We thank the CESGA team and specifically \'{A}lvaro Costas for the support in integrating Maestro and collecting benchmarks used to generate plots in this report. We thank the LRZ and specifically Jorge Echavarria for the support in developing and integrating the Maestro QDMI device.

\bibliographystyle{IEEEtran}

\begin{thebibliography}{10}
\providecommand{\url}[1]{#1}
\csname url@samestyle\endcsname
\providecommand{\newblock}{\relax}
\providecommand{\bibinfo}[2]{#2}
\providecommand{\BIBentrySTDinterwordspacing}{\spaceskip=0pt\relax}
\providecommand{\BIBentryALTinterwordstretchfactor}{4}
\providecommand{\BIBentryALTinterwordspacing}{\spaceskip=\fontdimen2\font plus
\BIBentryALTinterwordstretchfactor\fontdimen3\font minus \fontdimen4\font\relax}
\providecommand{\BIBforeignlanguage}[2]{{%
\expandafter\ifx\csname l@#1\endcsname\relax
\typeout{** WARNING: IEEEtran.bst: No hyphenation pattern has been}%
\typeout{** loaded for the language `#1'. Using the pattern for}%
\typeout{** the default language instead.}%
\else
\language=\csname l@#1\endcsname
\fi
#2}}
\providecommand{\BIBdecl}{\relax}
\BIBdecl

\bibitem{qiskit}
H.~Abraham and I.~Q.~D. Team, ``{Qiskit}: An open-source framework for quantum computing,'' \url{https://qiskit.org}, 2019, accessed: 2025-05-25.

\bibitem{cirq}
G.~Q.~A. Team, ``{Cirq}: A python framework for creating, editing, and invoking noisy intermediate scale quantum (nisq) circuits,'' \url{https://quantumai.google/cirq}, 2021, accessed: 2025-05-25.

\bibitem{projectq}
D.~S. Steiger, T.~H{\"a}ner, and M.~Troyer, ``Projectq: An open source software framework for quantum computing,'' \emph{Quantum}, vol.~2, p.~49, 2018.

\bibitem{cuquantum}
NVIDIA, ``{cuQuantum SDK},'' \url{https://developer.nvidia.com/cuquantum-sdk}, 2022, accessed: 2025-05-25.

\bibitem{quest}
T.~Jones, D.~W.~M. Brown, I.~Bush, and S.~C. Benjamin, ``Quest and high performance simulation of quantum computers,'' \emph{Scientific Reports}, vol.~9, no.~1, p. 10736, 2019.

\bibitem{maestro_repo}
\BIBentryALTinterwordspacing
A.~Roman and S.~DiAdamo, ``{Maestro: Multi-Backend Quantum Simulation Orchestration Engine},'' Nov. 2025. [Online]. Available: \url{https://github.com/QoroQuantum/maestro}
\BIBentrySTDinterwordspacing

\bibitem{viamontes2009quantum}
G.~F. Viamontes, I.~L. Markov, and J.~P. Hayes, \emph{Quantum circuit simulation}.\hskip 1em plus 0.5em minus 0.4em\relax Springer, 2009.

\bibitem{vidal2003efficient}
G.~Vidal, ``Efficient classical simulation of slightly entangled quantum computations,'' \emph{Physical review letters}, vol.~91, no.~14, p. 147902, 2003.

\bibitem{schollwock2011density}
U.~Schollw{\"o}ck, ``The density-matrix renormalization group in the age of matrix product states,'' \emph{Annals of physics}, vol. 326, no.~1, pp. 96--192, 2011.

\bibitem{orus2014practical}
R.~Or{\'u}s, ``A practical introduction to tensor networks: Matrix product states and projected entangled pair states,'' \emph{Annals of physics}, vol. 349, pp. 117--158, 2014.

\bibitem{qiskit2024}
A.~Javadi-Abhari, M.~Treinish, K.~Krsulich, C.~J. Wood, J.~Lishman, J.~Gacon, S.~Martiel, P.~D. Nation, L.~S. Bishop, A.~W. Cross, B.~R. Johnson, and J.~M. Gambetta, ``Quantum computing with {Q}iskit,'' 2024.

\bibitem{bayraktar2023cuquantum}
H.~Bayraktar, A.~Charara, D.~Clark, S.~Cohen, T.~Costa, Y.-L.~L. Fang, Y.~Gao, J.~Guan, J.~Gunnels, A.~Haidar \emph{et~al.}, ``cuquantum sdk: A high-performance library for accelerating quantum science,'' in \emph{2023 IEEE International Conference on Quantum Computing and Engineering (QCE)}, vol.~1.\hskip 1em plus 0.5em minus 0.4em\relax IEEE, 2023, pp. 1050--1061.

\bibitem{faj2023quantum}
J.~Faj, I.~Peng, J.~Wahlgren, and S.~Markidis, ``Quantum computer simulations at warp speed: Assessing the impact of gpu acceleration,'' \emph{arXiv preprint arXiv:2307.14860}, 2023.

\bibitem{jamadagni2024benchmarking}
A.~Jamadagni, A.~M. L{\"a}uchli, and C.~Hempel, ``Benchmarking quantum computer simulation software packages: state vector simulators,'' \emph{arXiv preprint ArXiv:2401.09076}, 2024.

\bibitem{walker1974new}
A.~J. Walker, ``New fast method for generating discrete random numbers with arbitrary frequency distributions,'' \emph{Electronics Letters}, vol.~10, no.~8, pp. 127--128, 1974.

\bibitem{leonteva2025comparative}
A.~Leonteva, G.~Masella, M.~Outteryck, A.~P. Orioli, and S.~Whitlock, ``Comparative benchmarking of utility-scale quantum emulators,'' \emph{arXiv preprint arXiv:2504.14027}, 2025.

\bibitem{proctor2022establishing}
T.~Proctor, S.~Seritan, E.~Nielsen, K.~Rudinger, K.~Young, R.~Blume-Kohout, and M.~Sarovar, ``Establishing trust in quantum computations,'' \emph{arXiv preprint arXiv:2204.07568}, 2022.

\bibitem{proctor2022measuring}
T.~Proctor, K.~Rudinger, K.~Young, E.~Nielsen, and R.~Blume-Kohout, ``Measuring the capabilities of quantum computers,'' \emph{Nature Physics}, vol.~18, no.~1, pp. 75--79, 2022.

\bibitem{AndresMartinez_Heunen2019}
\BIBentryALTinterwordspacing
P.~Andrés-Martínez and C.~Heunen, ``Automated distribution of quantum circuits via hypergraph partitioning,'' \emph{Physical Review A}, vol. 100, no.~3, Sep. 2019. [Online]. Available: \url{http://dx.doi.org/10.1103/PhysRevA.100.032308}
\BIBentrySTDinterwordspacing

\bibitem{kaur2025optimizedquantumcircuitpartitioning}
\BIBentryALTinterwordspacing
E.~Kaur, H.~Shapourian, J.~Zhao, M.~Kilzer, R.~Kompella, and R.~Nejabati, ``Optimized quantum circuit partitioning across multiple quantum processors,'' 2025. [Online]. Available: \url{https://arxiv.org/abs/2501.14947}
\BIBentrySTDinterwordspacing

\bibitem{vazquez2025cunqa}
J.~V{\'a}zquez-P{\'e}rez, D.~Exp{\'o}sito-Pati{\~n}o, M.~Losada, {\'A}.~Carballido, A.~G{\'o}mez, and T.~F. Pena, ``Cunqa: a distributed quantum computing emulator for hpc,'' \emph{arXiv preprint arXiv:2511.05209}, 2025.

\bibitem{wille2024qdmi}
R.~Wille, L.~Schmid, Y.~Stade, J.~Echavarria, M.~Schulz, L.~Schulz, and L.~Burgholzer, ``Qdmi-quantum device management interface: Hardware-software interface for the munich quantum software stack,'' in \emph{2024 IEEE International Conference on Quantum Computing and Engineering (QCE)}, vol.~2.\hskip 1em plus 0.5em minus 0.4em\relax IEEE, 2024, pp. 573--574.

\end{thebibliography}

\end{document}